\documentclass[prd,aps,twocolumn,amssymb,preprintnumbers,nofootinbib,%
amsmath]{revtex4}
\usepackage{hyperref}
\usepackage{epsfig}
\begin{document}
\title{Heavy cosmic strings}
\author{M.~Donaire}
\affiliation{DAMTP, CMS, University of Cambridge,
Wilberforce Road,
Cambridge CB3 0WA, United Kingdom
}
\author{A.~Rajantie}
\affiliation{Theoretical Physics, Blackett Laboratory, Imperial
College London, London SW7 2AZ, United Kingdom}

\date{21 February 2006}

\begin{abstract}
We argue that cosmic strings
with high winding numbers generally form in first order gauge 
symmetry breaking phase transitions, 
and we demonstrate this using computer simulations. 
These strings are heavier than single-winding strings
and therefore more easily observable.
Their cosmological evolution may also be very different.
\end{abstract}

\preprint{DAMTP-2005-76}

\maketitle

\section{Introduction}

Interest in cosmic strings~\cite{Kibble,VilShel} has recently been
resparked by new observational and theoretical findings.
It now seems that cosmic strings are
relatively generic prediction of superstring
theory~\cite{Copeland}, which is currently thought to be the most
promising candidate for a theory of quantum gravity. They also appear to be
unavoidable in grand unification based on
field theory, if the same theory is to explain
cosmological inflation as well as unification of elementary
particle interactions~\cite{Jeannerot}. Discovery of
cosmic strings would therefore open up an observational window 
to the very early universe, to extremely high energy physics and
possibly also to quantum gravity. A pair of apparently 
identical galaxies at redshift $z\approx 0.46$ known as
CSL-1~\cite{Sazhin1} caused a great deal of excitement because it seemed 
to bear the hallmarks of gravitational lensing by a string. Even though
pictures taken by the Hubble Space Telescope showed that it is, in fact,
nothing but two separate galaxies~\cite{Sazhin4}, it led to a large amount of 
theoretical work and significant advances in the field~\cite{KibbleDavis}.

In 1980's, cosmic strings were seen as a potential explanation for
the origin of structure in the universe~\cite{VilShel}, and
therefore their observational signatures have been studied in
detail. The strength of these signatures depends on the
dimensionless number $G\mu$, in which $G$ is Newton's
gravitational constant and $\mu$ is the tension of the
string, defined as the energy per unit length. So far,
cosmic string searches have not found anything, which sets an upper
bound on the tension. The strongest bounds arise from the cosmic
microwave background \cite{Wyman} and timing of
pulsars~\cite{Vachaspati,Lommen}; both give $G\mu\lesssim
\mbox{few} \times 10^{-7}$. The natural value of $G\mu$ in
superstring theory models seems to be significantly lower, around
$10^{-10}\ldots 10^{-9}$~\cite{Copeland}.
While these strings would be observable with planned gravitational wave experiments
such as LISA,~\cite{Damour} they are clearly out of reach of our current observations.

However, theories with cosmic strings predict rather generally a spectrum of different
string tensions.
This is the case in field theory models in the ``Type-I'' regime, where strings with any integer
winding number are stable, as well as superstring models in which there are
stable bound states of D and F-strings known as $(p,q)$-strings~\cite{Copeland}.
As was mentioned in Ref.~\cite{Tye},
cosmic string searches are generally constraining a somehow averaged value
of the tension. To be specific, let us denote the tension of a single-winding string
by $\mu_0$, and the ratio of the tension of a given string to this
by $x=\mu/\mu_0$. Because the amplitude of gravitational waves
behaves as $\Omega_g\propto (G\mu)^{2/3}$ in the relevant
range~\cite{Damour}, the pulsar bound actually constrains $\overline{x^{2/3}}
G\mu_0$ rather than the tension $G\mu_0$ of an elementary single-winding string. 
This means that if multiple-winding strings exist, these searches can be much more 
sensitive than they appear.

The cosmological evolution of a network of strings with multiple windings
is an open problem, although some studies have recently been done.
For instance, Ref.~\cite{Tye} treats the interactions of strings as local events 
and ignores their extended nature. Field theory simulations of multiple-winding cosmic string
interactions~\cite{Laguna,Bettencourt} have shown that interactions
typically lead to lower winding numbers.
For multiple-winding strings to exist, they
must have been formed with high winding numbers in the first place.
The aim of this paper is to show that this can indeed happen.

\section{Setup}
We will focus on cosmic string formation in
a finite-temperature phase transition in the Abelian Higgs model.
Realistic grand unified theories should behave qualitatively in the same
way, and experience with superstring theory models seems to indicate that
the same general principles would apply also in that case~\cite{DvaliVilenkin}.
The model consists of two fields, the complex Higgs scalar $\phi$
and the gauge field $A_\mu$ associated with an U(1) gauge symmetry.
The Lagrangian is
\begin{equation}
{\cal L}=
-\frac{1}{4}F_{\mu\nu}F^{\mu\nu}
+(D_\mu\phi)^*D^\mu\phi-\lambda\left(\phi^*\phi-\eta^2\right)^2,
\label{equ:lagr}
\end{equation}
where we have used the field strength tensor $F_{\mu\nu}=
\partial_\mu A_\nu-\partial_\nu A_\mu$ and the covariant derivative
$D_\mu=\partial_\mu+ieA_\mu$. The model has two dimensionless parameters,
the gauge coupling constant $e$ and the Higgs self-coupling $\lambda$,
and the dimensionful scale $\eta$, which corresponds to the vacuum expectation
value of $\phi$.
The corresponding equations of motion in Minkowski space are
\begin{eqnarray}
\partial_\mu F^{\mu\nu}&=&-2e{\rm Im}\phi^* D^\nu\phi,\nonumber\\
D_\mu D^\mu \phi &=& -2\lambda \left(\phi^*\phi -\eta^2\right)\phi.
\label{equ:eom}
\end{eqnarray}

To model cosmic string formation in the early universe~\cite{Kibble,OwnReview},
we study our model at a non-zero temperature.
It has two phases: the high-temperature symmetric (Coulomb) phase
characterized by a massless photon field,
and the low-temperature broken (Higgs) phase
which is essentially a relativistic version of a superconducting phase.

As in superconductors,
the ``magnetic field'' $\vec{B}=\vec\nabla\times\vec{A}$
cannot exist in the broken phase. Instead, it gets
confined
into Nielsen-Olesen vortex lines, which
are our cosmic strings.
The flux $\Phi$ carried by a string is an
integer multiple of the flux quantum $\Phi_0=2\pi/e$, and this
defines the winding number $N_W=\Phi/\Phi_0$ of the string.
At zero temperature, strings with $N_W>1$ are stable if
$e^2/\lambda>1$, and at high temperatures one requires
$e^2/\lambda\gtrsim 1.7$~\cite{Mo}. 
In this so-called Type I regime, the transition is of first order.

We investigate the model in the Type I regime. The system is initially in
thermal equilibrium in the symmetric phase, and the temperature is
then gradually decreased. When the temperature falls below a
certain value, the symmetric phase becomes metastable. Bubbles of
broken phase nucleate, grow and
eventually fill the whole space. In our analytical calculations, we assume that the transition is
strongly first order so that by this time their typical size is 
much larger than the microscopic
scales such as the thickness of bubble walls or their radius at nucleation.

\section{String formation}
It is often assumed that cosmic strings
are formed by the Kibble mechanism~\cite{Kibble}, even though it is
strictly speaking only valid in the global limit $e=0$.
According to this picture, the complex phase of $\phi$ has a
random but constant value in each bubble. When two bubbles collide, 
the phase angle
takes the shortest way to interpolate between the two values. 
Whenever three bubbles meet, it is possible that a
hole of symmetric phase is left between them. There
is then a probability of $1/4$ that, when going from bubble 1 to
bubble 2 to bubble 3, the phase angle changes by $2\pi$. If this happens, 
a string is formed when the
hole eventually closes. The number density of
strings per unit cross-sectional area is therefore roughly
$1/R^2$, where $R$ is the typical bubble size.
Note that only strings with $N_W=1$ can form. Even
$N_W=2$ would require a simultaneous collision of at
least five different bubbles at the same point, and since we are assuming that the wall thickness is negligible, this never happens.

Away from the global limit, the Kibble mechanism does not
fully describe string formation~\cite{OwnReview,Manuel}.
Strings are also produced by thermal fluctuations
of the magnetic field, which become trapped in regions of symmetric 
phase~\cite{KibbleVilenkin,OwnReview}.
As we will show, this mechanism can form strings with $N_W>1$.

In thermal equilibrium the state of the system is given by an ensemble of 
configurations with
a Boltzmannian probability
distribution, $p\propto\exp(-E/T)$, where $E=\int d^3x \rho$ is the total 
energy of the system.
In the
temporal gauge ($A_0=0$), the energy density is
\begin{equation}
\rho=\frac{1}{2}\left(\vec{E}^2+\vec{B}^2\right) +\dot\phi^*\dot\phi
+\left|\vec{D}\phi\right|^2
+\lambda\left(\phi^*\phi-\eta^2\right)^2,
\label{equ:hamiltonian}
\end{equation}
where $\vec{E}=-\partial_0\vec{A}$ and $\vec{B}=\vec\nabla\times\vec{A}$.
Most importantly for us, $\vec{B}$ has approximately Gaussian thermal 
fluctuations.

\begin{table}
\begin{tabular}{l|c|r}
lattice spacing & $\delta x$ & 1\cr
time step & $\delta t$ & 0.05\cr
gauge coupling & $e$ & 1\cr
scalar coupling & $\lambda$ & 0.05\cr
Higgs vev & $\eta$ & $\sqrt{5}$ \cr
initial temperature & $T$ & 0.78
\end{tabular}

\caption{\label{table:params}
Parameter values used in the simulations.
}
\end{table}

Magnetic flux may get swatted between two colliding bubbles~\cite{Manuel},
but in this paper we will focus on three-bubble collisions, because 
they can produce higher winding numbers~\cite{KibbleVilenkin}.
When the three bubbles coalesce leaving a hole of symmetric phase 
between them, 
any flux that was there initially will remain trapped in the hole because flux
lines cannot penetrate the broken phase. When the hole closes, the flux 
trapped in it forms a string. 

The winding number of the string is determined by the trapped flux,
with a possible contribution of $\pm 1$ from the Kibble mechanism, which
we will ignore from now on. We must therefore estimate
how much flux there was in the region before it became trapped.
At that time, it was still in thermal equilibrium with its surroundings,
and therefore the flux is determined by the thermal equilibrium distribution.
We can calculate its typical value using the saddle point method.
In the symmetric phase, where $\phi=0$, 
the energy density is proportional to $\vec{B}^2$.
This means that the minimal energy $E_{\rm
min}(\Phi)$ for a configuration with given magnetic flux $\Phi$
through the region is proportional to $\Phi^2$. 
For dimensional reasons it has to be of the form
$E_{\rm min}(\Phi)=C\Phi^2/R,$
where $R$ is the typical bubble size and $C$ is a dimensionless number
of order one.

The typical value $\Phi_{\rm rms}$ of the flux,
given by the square root of the variance of the probability
distribution $p(\Phi)\propto\exp(-E_{min}(\Phi)/T)$,
is
$\Phi_{\rm rms}\approx \sqrt{RT}$.
When the hole between the bubbles eventually closes, all this flux turns
into a string with typical winding number
\begin{equation}
\label{equ:prediction}
N_W\approx \Phi_{\rm rms}/\Phi_0\approx 
(e/2\pi)\sqrt{RT}\approx (e/2\pi)(v/\Gamma)^{1/8}T^{1/2}.
\end{equation}
Here we have used dimensional analysis to express $R$
in terms of the bubble wall velocity $v$
and the bubble nucleation rate $\Gamma$
as $R\approx(v/\Gamma)^{1/4}$.

Because $R$ is much greater than the thermal
wavelength $1/T$, $N_W$ can be large. 
If we, for instance, assume that $R$ is comparable to
the Hubble radius $1/H$, we find
\begin{equation}
N_W\approx \frac{e}{2\pi g_*^{1/4}}\sqrt{\frac{m_{\rm Pl}}{T}}\approx
\frac{e}{2\pi (g_*G\mu_0)^{1/4}},
\end{equation}
where $g_*$ is the effective number of thermal degrees of freedom.
If the tension $\mu_0$
of a single-winding string is low, the typical winding number
can therefore be high. In particular, this means that the typical observed
tension $\mu\approx N_W\mu_0$ scales as $(G\mu_0)^{3/4}$
rather than $G\mu_0$, making the strings more easily detectable.
For $e=1$ and $g_*\approx 100$, this would strengthen the pulsar 
bound by about a factor of two.

\begin{figure}
\begin{center}
\epsfig{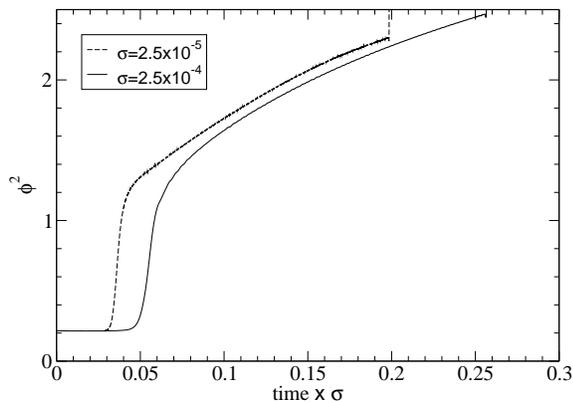}
\end{center}
\caption{\label{fig:phi2}
Time evolution of $\langle\phi^2\rangle$ for different cooling rates.
The time axis
has been rescaled by the cooling rate, so that it is in one-to-one
correspondence with the
effective temperature.
}
\end{figure}

\begin{figure}
\begin{center}
\begin{tabular}{ll}
\epsfig{file=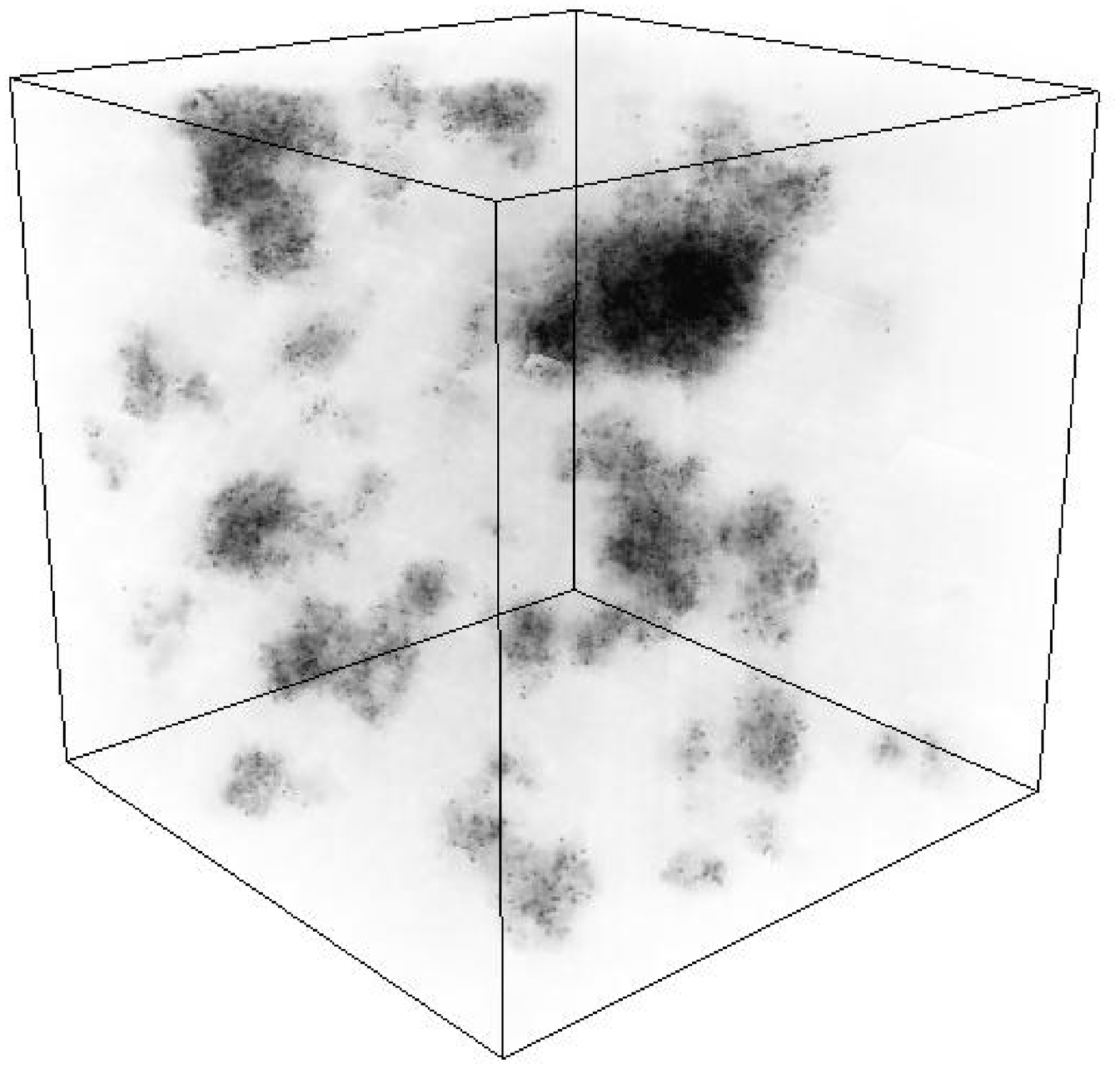,width=4.5cm}&
\epsfig{file=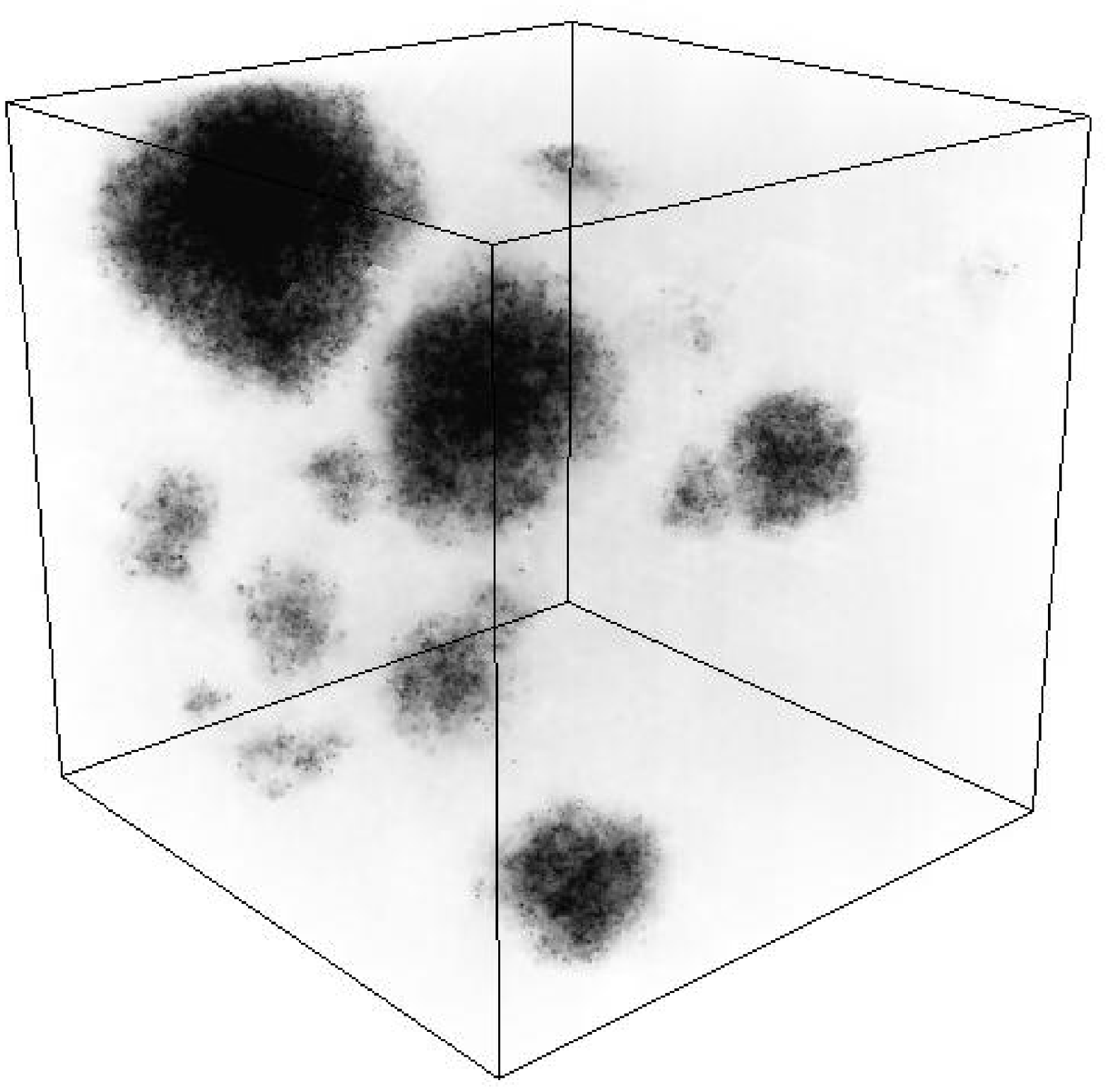,width=4.5cm}\\[-4mm]
(a)&(b)
\end{tabular}
\end{center}
\caption{
\label{fig:bubbles}
Bubbles of broken phase at the time when $\langle\phi^2\rangle=0.33$
in the simulations with cooling rates
(a) $\sigma=2.5\times 10^{-4}$ and (b) $\sigma=2.5\times 10^{-5}$.
With lower cooling rate, the bubbles are larger and there are fewer of them,
because the nucleation rate was lower at the time of their nucleation.}

\end{figure}

\begin{figure}
\vspace*{-5mm}
\begin{center}
\epsfig{file=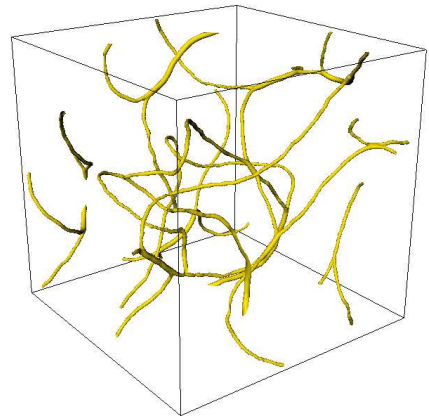,width=7.5cm}\\[-4mm]
$\sigma=2.5\times 10^{-4}$\\
\epsfig{file=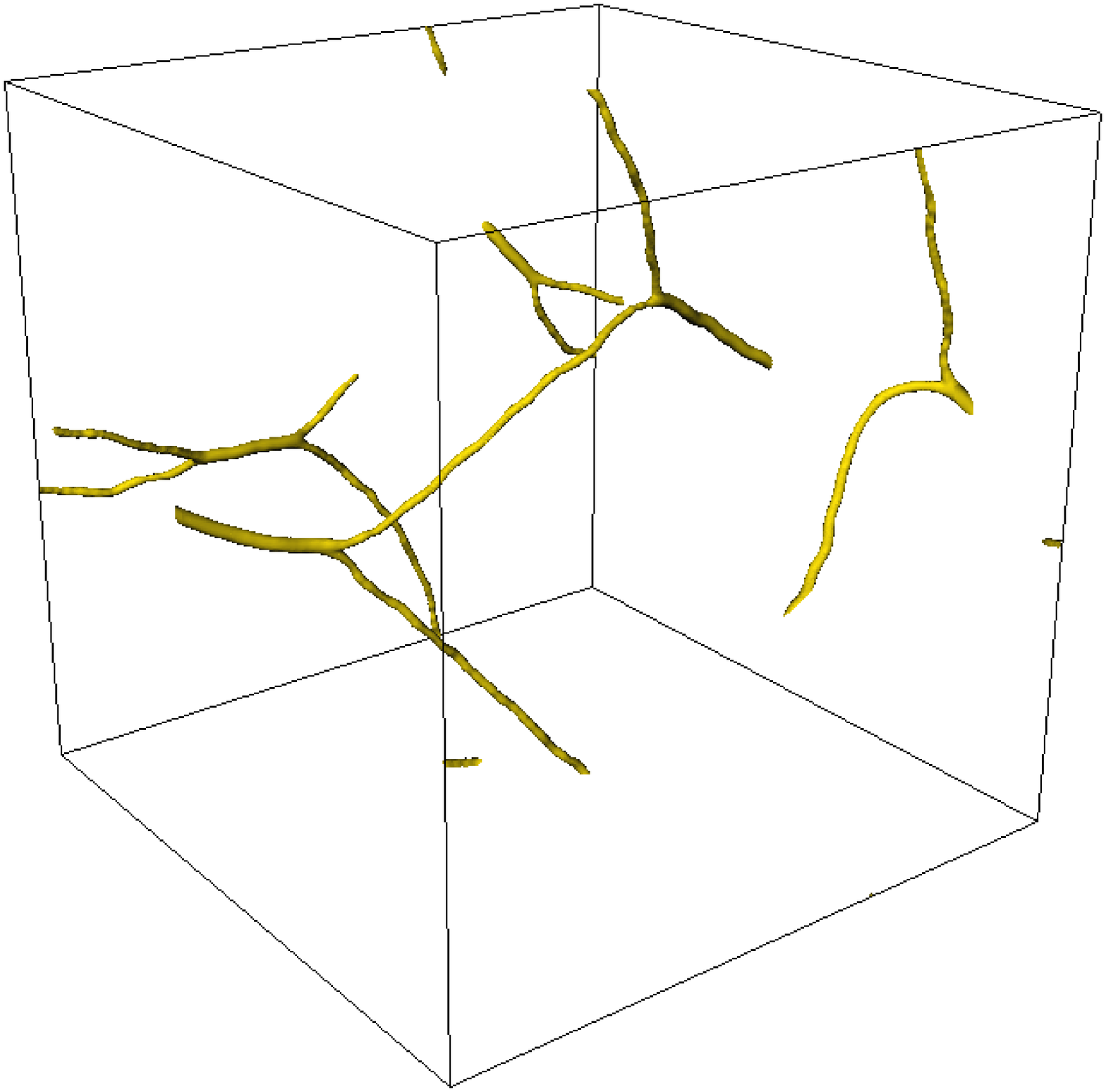,width=7.5cm}\\[-1mm]
$\sigma=2.5\times 10^{-5}$\\[-2mm]
\end{center}
\caption{
\label{fig:strings}
String networks formed in simulations with two different cooling
rates $\sigma$.
In both cases, we have used
a short period of gradient flow to remove short-distance thermal noise,
and plotted isosurfaces of $|\phi|$.
We have made movies of the full time evolution available at
\href{http://www.imperial.ac.uk/people/a.rajantie/research/heavystrings/}%
{http://www.imperial.ac.uk/people/a.rajantie/research/heavystrings/}.
}
\end{figure}

\section{Simulations}
To test this theoretical picture, we carried out numerical
simulations of first-order phase transitions in the Abelian Higgs model.
We discretized the theory on a $256^3$
lattice in the standard
way \cite{Moriarty} using the non-compact formulation (see, for instance,
Ref.~\cite{latticebook}).
The values of all parameters are given in Table \ref{table:params}.
We first generated thermal initial conditions by
evolving the system in time using the classical equations of motion
(\ref{equ:eom}) and
generating random values at random intervals for those components of the time
derivatives $\dot\phi$ and $\vec{E}=-\partial_0\vec{A}$
that do not appear in Gauss's law
$\vec\nabla\cdot\vec{E}=2e{\rm Im}\phi^*\dot\phi$.

After $512$ units of time and $64$ randomization steps, we were satisfied that
the state of the system was in equilibrium. After that, we added a
small damping term $\sigma$
to the equations of
motion (\ref{equ:eom}), and evolved the system in time without
any random noise.
The damping term was implemented in such a way that it only affected the components of the
field that do not appear in Gauss's law.
It cools the system down as $T_{\rm eff}\propto \exp(-\sigma t)$
and serves two purposes: It causes the phase transition to start earlier because the 
bubble nucleation rate depends on the temperature, and it removes the latent heat released by the growing bubbles. Without damping, the system would end up in a state of two coexisting phases.
We ran the simulation with two different cooling rates
$\sigma=2.5\times 10^{-4}$ and $2.5\times 10^{-5}$.

The time evolution of the variance $\langle\phi^2\rangle$ of the scalar field
is shown in Fig.~\ref{fig:phi2}, and consists of three stages:
First, the system is in the symmetric phase, with a small and almost constant
$\langle\phi^2\rangle$. Then, at around $t\approx 0.05/\sigma$,
there is a rapid transition to the broken phase, after which the
system equilibrates, apart from
a small number of strings that were formed in the transition.
When the cooling is slower, the transition takes places at a higher temperature.
Because the nucleation rate is then lower,
fewer bubbles are nucleated and they can become larger.
To confirm this, we have plotted in Fig.~\ref{fig:bubbles} the regions that
are in the broken phase at the instant when $\langle\phi^2\rangle=0.33$.

Pictures of the string networks produced in the two runs
are shown in 
Fig.~\ref{fig:strings}.
As expected, slower cooling
produced fewer strings. 
The ratio of the total string length between the runs is $2.3$, indicating
a ratio of typical bubble sizes of $1.5$. 

In both cases, one can see several
junctions where two strings merge into an heavier winding-2 string.
This would not happen in the Kibble mechanism. The length ratio of windings
2 and 1 is higher in the slower transition (0.18 vs 0.15), supporting
our theory. However, because of the low statistics and finite-size
effects, this cannot be directly compared with the prediction in 
Eq.~(\ref{equ:prediction}), which furthermore assumes that $N_W\gg 1$.
This condition is
not satisfied in this simulation or indeed in any similar simulation unless
one can use several orders of magnitude larger lattice sizes.

\section{Conclusions}
We have shown that thermal first-order phase
transitions can produce strings with high winding numbers and
predicted the typical winding in Eq.~(\ref{equ:prediction}).
Our numerical simulations confirm the formation of strings with $N_W>1$.
Quantitative tests are not possible with simulations of this type because
of the computing power they would require, but other numerical tests will
be reported in a future publication~\cite{Manuel}.
It is not known how a string network with different winding numbers evolves,
but if the heavy strings survive until today, they will be much more easily observable
than ordinary single-winding strings. The current observational 
upper bounds would therefore become much stronger.
This will, however, require a large scale simulation of the later evolution of the string network
and is beyond the scope of this paper.
It would also
be interesting to extend this work to brane collisions in
superstring theory models to find out whether heavy strings are also 
produced in that case by a similar mechanism. 

\acknowledgments
This research was conducted in cooperation with SGI/Intel
utilising the Altix 3700 supercomputer and was supported by 
the ESF Coslab programme. 
MD was supported by EPRSC and 
The Cambridge European Trust, and AR by Churchill College, Cambridge.


\begin{thebibliography}{100}
\bibitem{Kibble}
  T.~W.~B.~Kibble,
  J.\ Phys.\ A {\bf 9}, 1387 (1976).

\bibitem{VilShel}
A.~Vilenkin and E.~P.~S.~Shellard,
``Cosmic Strings and Other Topological Defects''
(Cambridge University Press, Cambridge, 1994).

\bibitem{Copeland}
  E.~J.~Copeland, R.~C.~Myers and J.~Polchinski,
  JHEP {\bf 0406}, 013 (2004).

\bibitem{Jeannerot}
  R.~Jeannerot, J.~Rocher and M.~Sakellariadou,
  Phys.\ Rev.\ D {\bf 68}, 103514 (2003).

\bibitem{Sazhin1}
  M.~Sazhin {\it et al.},
  Mon.\ Not.\ Roy.\ Astron.\ Soc.\  {\bf 343}, 353 (2003).

\bibitem{Sazhin4}
  M.~V.~Sazhin, M.~Capaccioli, G.~Longo, M.~Paolillo and O.~S.~Khovanskaya,
  arXiv:astro-ph/0601494.
  
\bibitem{KibbleDavis}
  A.~C.~Davis and T.~W.~B.~Kibble,
  Contemp.\ Phys.\  {\bf 46}, 313 (2005)
  [arXiv:hep-th/0505050].

\bibitem{Wyman}
  M.~Wyman, L.~Pogosian and I.~Wasserman,
  Phys.\ Rev.\ D {\bf 72}, 023513 (2005).

\bibitem{Vachaspati}
  T.~Vachaspati and A.~Vilenkin,
  Phys.\ Rev.\ D {\bf 31}, 3052 (1985).

\bibitem{Lommen}
  A.~N.~Lommen,
  arXiv:astro-ph/0208572.

\bibitem{Damour}
  T.~Damour and A.~Vilenkin,
  Phys.\ Rev.\ D {\bf 71}, 063510 (2005).

\bibitem{Tye}
  S.~H.~Tye, I.~Wasserman and M.~Wyman,
  Phys.\ Rev.\ D {\bf 71}, 103508 (2005)
  [Erratum-ibid.\ D {\bf 71}, 129906 (2005)].

\bibitem{Laguna}
  P.~Laguna and R.~A.~Matzner,
  Phys.\ Rev.\ Lett.\  {\bf 62}, 1948 (1989).

\bibitem{Bettencourt}
  L.~M.~A.~Bettencourt, P.~Laguna and R.~A.~Matzner,
  Phys.\ Rev.\ Lett.\  {\bf 78}, 2066 (1997).


\bibitem{DvaliVilenkin}
  G.~Dvali and A.~Vilenkin,
  JCAP {\bf 0403}, 010 (2004).

\bibitem{OwnReview}
  A.~Rajantie,
  Int.\ J.\ Mod.\ Phys.\ A {\bf 17}, 1 (2002).

\bibitem{Mo}
  S.~Mo, J.~Hove and A.~Sudbo,
  Phys.\ Rev.\ B {\bf 65}, 104501 (2002).

\bibitem{Manuel}
M.~Donaire, in preparation.

\bibitem{KibbleVilenkin}
  T.~W.~B.~Kibble and A.~Vilenkin,
  Phys.\ Rev.\ D {\bf 52}, 679 (1995).

\bibitem{Moriarty}
  K.~J.~M.~Moriarty, E.~Myers and C.~Rebbi,
  Phys.\ Lett.\ B {\bf 207}, 411 (1988).

\bibitem{latticebook}
J.~Smit, ``Introduction to Quantum Fields on a Lattice''
(Cambridge University Press, Cambridge, 2002).

\end{thebibliography}
\end{document}